\documentclass[fleqn,11pt]{wlscirep}

\usepackage{amsmath}

\usepackage{mathtools}

\usepackage{amssymb}
\usepackage{amsfonts}

\usepackage{relsize}

\usepackage{placeins}
\usepackage{amssymb}
\usepackage{hyperref}
\usepackage[none]{hyphenat}
\usepackage{graphicx,subcaption}
\usepackage{float}
\usepackage{todonotes}
\usepackage{layouts}
\usepackage{tikz}
\usepackage[outdir=./]{epstopdf}

\newcommand{\mean}[1]{\left<{#1}\right>}
\newcommand{\ER}{Erd\H{o}s-R\'enyi }
\newcommand{\ERR}{\emph{ER}}
\newcommand{\RGG}{\emph{RG}}

\begin{document}
\title{Spatial effects in meta-foodwebs} 
\author[1*]{Edmund Barter}
\author[1]{Thilo Gross}
\affil[1]{University of Bristol, Department of Engineering Mathematics, Bristol, UK}
\affil[*]{edmund.barter@bristol.ac.uk}
\begin{abstract}
In ecology it is widely recognised that many landscapes comprise a network of discrete patches of habitat. The species that inhabit the patches interact with each other through a foodweb, the network of feeding interactions. The meta-foodweb model proposed by Pillai et al. combines the feeding relationships at each patch with the dispersal of species between patches, such that the whole system is represented by a network of networks. Previous work on meta-foodwebs has focussed on landscape networks that do not have an explicit spatial embedding, but in real landscapes the patches are usually distributed in space. Here we compare the dispersal of a meta-foodweb on \ER networks, that do not have a spatial embedding, and random geometric networks, that do have a spatial embedding. We found that local structure and large network distances in spatially embedded networks, lead to meso-scale patterns of patch occupation by both specialist and omnivorous species. In particular, we found that spatial separations make the coexistence of competing species more likely. Our results highlight the effects of spatial embeddings for meta-foodweb models, and the need for new analytical approaches to them.
\end{abstract}
\maketitle
\section{Introduction}

\begin{figure}[t]
\centering
\includegraphics{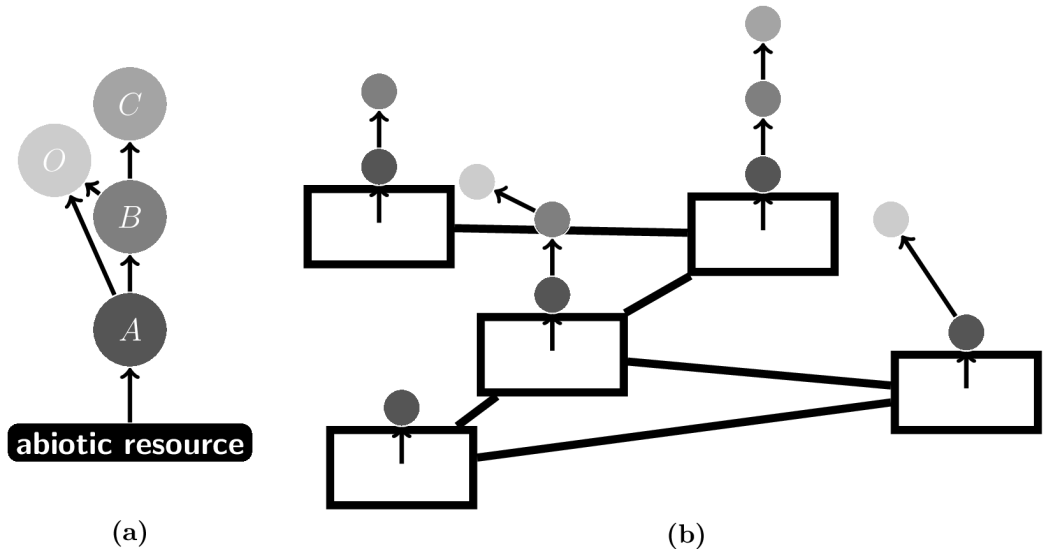}
\caption{\label{fig1} Meta-foodwebs as networks of networks. Panel (a) shows a simple meta-foodweb of four species. Each node represents a species and the directed links are from prey to predators in a feeding relationship. Species $A$ is a primary producer, while species $B$ and $C$ are specialist consumers. Species $O$ is an omnivore, which can feed on multiple trophic levels. Panel (b) shows a patch network. Rectangles represent each patch and links are between patches which species can disperse between. Each patch is occupied by a local food chain. The local networks change in time due to colonization and extinction events. }
\end{figure}

Foodwebs, the networks of trophic (feeding) interactions among a community of species, are among the paradigmatic examples of complex networks. Their composition and dynamics have been studied extensively.\cite{May1972,McCann1998,McCann2000,Neutel2002,Montoya2006,Gross2009,Lafferty2015} In nature, communities are often not isolated but are embedded in a complex structured environment that consists of distinct patches of habitat.\cite{Levins1969,Hanski1999} Depending on the system under consideration the patches may be lakes, islands, or actual patches of forest left in an agricultural landscape. In typical environments, the communities at many similar patches interact through the dispersal of individuals between neighbouring patches. The aggregations of the foodwebs at patches related by a complex spatial network are called meta-foodwebs.\cite{Warfe2013,Gramlich2016} These systems, comprising local foodwebs joined to each other by links between patches, can be represented by a network of networks \cite{Kivela2014}(Fig~\ref{fig1}b).

The study of spatial interactions has a long history in Ecology. For instance in explaining the global coexistence of similar competitors\cite{MacArthur1967,Harrison1991} and the survival of multiple species on the same limiting resource.\cite{Levins1971,Levin1974,Nee1992} Studies in this area often account for the presence or absence of the species at each patch as a binary variable.\cite{Hanski1991,Hanski1998} In these, so-called patch-dynamic, models the state of each patch changes in time due to local colonization and extinction events. 

Most patch-dynamic models focus on simple cases such as single populations or competitive interactions between similar species. \cite{Hanski1999a} However, recently Pillai~et~al.\cite{Pillai2009,Pillai2011} set out a framework that incorporates trophic interactions between species into patch-dynamic models. This meta-foodweb model considers complex foodwebs of many species with predator-prey and competitive interactions.

The meta-foodweb model of Pillai~et.~al. has been used to demonstrate that general spatial heterogeneity can increase stability of complex foodwebs.\cite{Pillai2011,Gravel2011} More specifically the number and distribution of links in the network of patches have been shown to have non-trivial effects on the distribution of foodwebs among the patches.\cite{Bohme2012,Barter2016} We have previously shown that species at different levels of a food chain may benefit from different distributions of patch degrees (number of links to other patches).\cite{Barter2016} Other previous work addressed the prominent question of whether, and under what conditions an omnivore predator can coexist with a specialist, who is a stronger competitor.\cite{Bohme2012} This demonstrated that an omnivore can persist only when the average number of links at each patch, the mean degree, is within a particular range. 

Previous theoretical works investigated meta-foodwebs where the underlying patch network was assumed to be an \ER random graph or configuration model network.\cite{Bohme2012,Barter2016} In such networks any given pair of patches has a fixed chance of interacting and therefore the networks typically have a small \emph{diameter}, a measure of the maximum distance between nodes.\cite{Newman2010} In these networks the shortest path (series of links) between any given pair of nodes is small \cite{Klee1981,Bollobas1981} and they are subsequently termed small-worlds. \cite{Watts1998}  By contrast, real world meta-foodwebs are constrained by geography and the individual's ability to travel between patches. Such a spatial embedding constrains the possible networks as  patches are only linked if they are close enough together for individuals to disperse between them. Geometric distances are translated in network distances and in the resulting network the shortest path between two patches can be relatively long.\cite{Banavar1999,Dall2002} 

In a small world network, a species can quickly disperse from any node to every other node. By contrast, in a large world pronounced geographical barriers, characterized by long network path lengths, may exists that impede rapid colonization of distant parts. While we will provide a more detailed analysis below it is intuitively conceivable that the large world nature of spatially embedded networks of patches creates spatial niches in which a species can survive with relatively little danger from competitors. Moreover, in small worlds the neighbours of any particular node tend to be a representative sample from the network. A node in a small world is thus exposed to colonization from the full range of communities that the system supports. By contrast, in large worlds the neighbours of a node are located in the same region of the network as the focal node, and most colonization will be from communities that are very similar to the one established in the focal node. This reinforcement of communities may further promote species persistence.

Here we investigate the effects of spatial nature of a patchy environment (i.e.~the large-worldishness) on the dispersal of foodwebs. We build our analysis on a comparison of non-spatial \ER networks\cite{erdds1959random,Gilbert1959,erdos1961evolution} and explicitly spatial, random geometric patch networks.\cite{Dall2002} We find two results: First, specialist consumers are less abundant (occupy fewer patches) on the spatial patch networks. Second, when the landscape is also occupied by a competitor, generalist consumers are more abundant on the spatial patch networks than on non-spatial patch networks. We conclude that these results are predominantly due to the larger distances between patches in the spatial networks.

\section{The Model}

We study a version of the model proposed by Pillai~et~al..\cite{Pillai2009} The model describes a set of species, each of which either occupies or is absent from each patch in a spatial network at each moment in time. Trophic interactions between the species are represented by a global meta-foodweb (Fig.~\ref{fig1}a). Following Pillai et al.\cite{Pillai2009} we assume each patch contains only a subset of the species of the global foodweb, and these comprise a local food chain (Fig.~\ref{fig1}b). The global meta-foodweb comprising all the species only becomes evident when an aggregation of the spatial system is considered.

The trophic interactions of the meta-foodweb have implications on the ability of species to occupy each patch. A species must be able to feed at every patch it occupies. Primary producers can occupy an empty patch, but all other species can only occupy patches where their prey is present. Furthermore a species cannot share a patch with another species competing for the same prey. Following Pillai et al.\cite{Pillai2009} we assume that specialist consumers outcompete their generalist counterparts for a particular prey all of the time. Hence, all local foodwebs are linear chains. 

\begin{figure*}[t]
\centering
\includegraphics{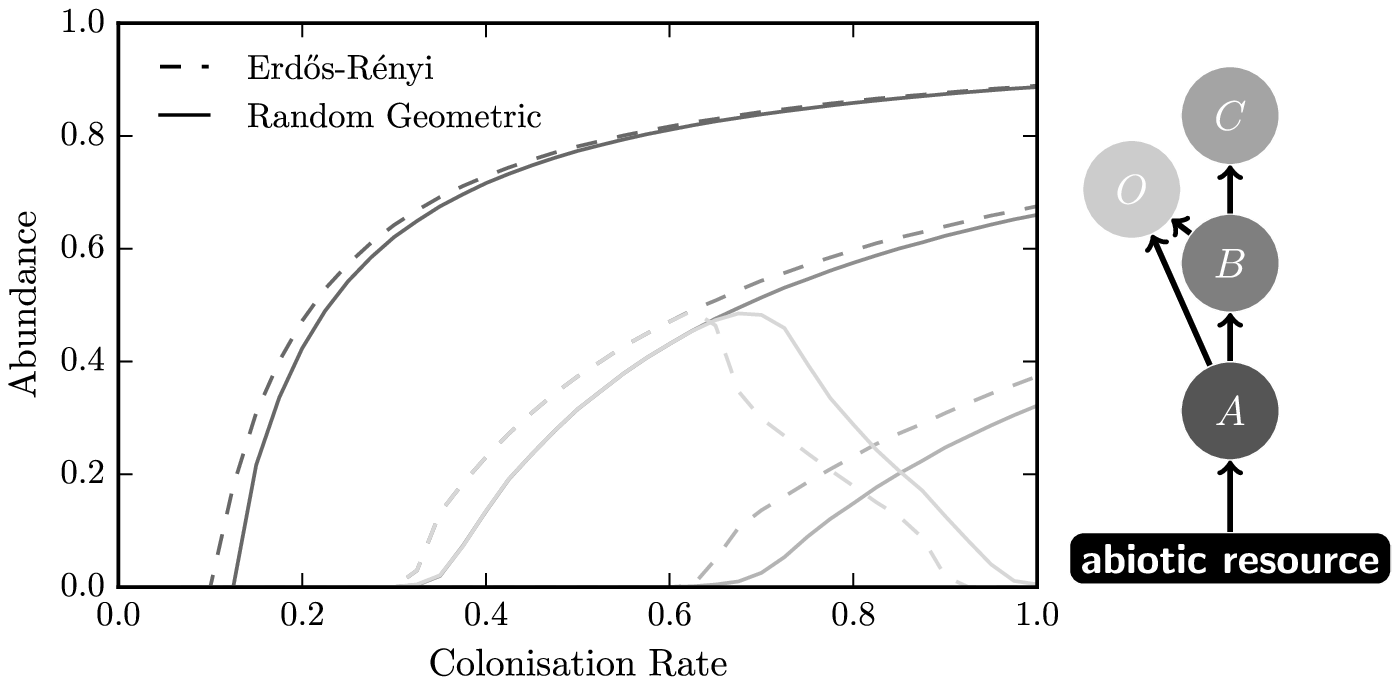}
\caption{Spatial structure affects species abundance. The abundance (mean occupied fraction) of the species in the meta-foodweb on \ER (dashed lines) and random geometric (solid) networks. The colours of the lines correspond to the colours of the species in the foodweb on the right. Most of the species are more abundant on \ER networks. The exception is the omnivore, which is more abundant on random geometric networks when coexisting with a specialist.}
\label{abundance_against_c}
\end{figure*}

A patch that contains at least one food source that can be utilized by a given species, and does not contain a superior competitor to that species, can be occupied by it and we say the patch is available to that species.

The species that occupy a particular patch change in time due to the colonization of the patch by species from neighbouring patches and the extinction of the local populations at the patch. When established on a patch, species $X$ colonises neighbouring patches that are available to it at the rate $c_X$. When established at a patch, species $X$ also goes extinct on that patch at the rate $e_X$. In the following we set $e_X=1$ (arbitrary units) and further assume that $c_X=c$ for all species in the meta-foodweb. 

At all times the local foodwebs must satisfy the restrictions imposed on species by trophic interactions. Therefore, when species $X$ is established on a patch where its only prey is species $Y$, and species $Y$ goes extinct on that patch, species $X$ will also go extinct, we call this process \emph{indirect extinction}. When a specialist predator, species $X$, colonises a patch that is occupied by an omnivore species $Y$, and species $X$ and species $Y$ share a food source,  then species $Y$ may no longer exist on that food source. This will lead to the extinction of species $Y$ on that patch, unless it can prey on species $X$, so has a food source it does not compete for. We call the extinction of a generalist by an arriving specialist \emph{driven extinction}.

To study the effects of space on species coexistence we consider dynamics on two different types of  patch networks. The first type of network are those from the \ER (\ERR{}) ensemble.\cite{erdds1959random,Gilbert1959} These networks are generated by taking a set of nodes and assigning a link between each pair with a probability $p=z/N$, where $N$ is the number of nodes and $z$ is the desired mean degree, i.e. the expectation value of the number of neighbours for a randomly chosen node.\cite{erdos1961evolution}

A central property of networks that determines how easy a particular network is to colonize is the degree distribution $p_k$, the probability distribution that a given node has $k$ links.\cite{Pastor-Satorras2001,Barter2016} For \ERR{} networks the degree distribution is poisson distributed, so $p_k= z^ke^{-z}/k!$. Because links are added randomly between nodes the \ERR{} is a non-spatial network and every node can be reached from every other node in a small number of steps, which scales as $\log(N)$.\cite{Bollobas1981}

The second type of network considered here are from the ensemble of $2$-dimensional random geometric (\RGG{}) networks .\cite{Penrose2003} These networks are generated by, first, randomly distributing $N$ nodes in a unit square with uniform distribution and then adding links between all pairs of nodes that are within a distance $r$ of each other. The degree distribution of \RGG{} networks is the same as that for \ERR{} networks, $p_k= z^ke^{-z}/k!$ where the mean degree is given by $z=\pi r^2 N$.  The spatial distances between points in \RGG{} networks are translated into network distances and therefore the shortest path between some pairs of nodes is large and scales with $\sqrt{N}/r$.\cite{Friedrich2013}

Although the \ERR{} and \RGG{} networks have the same degree distribution, they have different levels of degree correlations. In \ERR{} networks a link is equally likely between any pair of nodes, and so the degrees of neighbouring nodes are uncorrelated. In \RGG{} networks each node is at the centre of a circle with radius $r$ that contains all its neighbours, we call this its \emph{spatial neighbourhood}. A node's degree is the number of nodes in its spatial neighbourhood. Two nodes that have a link between them in the network must have overlapping spatial neighbourhoods. As the number of nodes in the intersection of their spatial neighbourhoods is the same, the degree of a node in a random geometric graph is correlated with the degree of its network neighbours.

We simulate the dispersal of the species in the meta-foodweb on networks from each ensemble using a Gillespie-type algorithm .\cite{Gillespie1976} By choosing ensembles with matching degree distributions we eliminate the effects on dispersal considered previously,~\cite{Bohme2012,Barter2016} and focus on the differences due to the difference in path lengths between the ensembles.  

For each species in the system there are two possible equilibrium states, either the species is extinct at the metapopulation level, i.e. does not occupy any patches, or the metapopulation is a constant size, i.e.~it does. After the simulation has reached this equilibrium for all species we record the time for which each patch was occupied by each possible configuration of the species over the rest of the simulation. From these states we calculate the fraction of time each species occupies each patch (the patch occupation), the sum of which over all patches we term the species abundance.  

\begin{figure*}[t]
\centering
\includegraphics{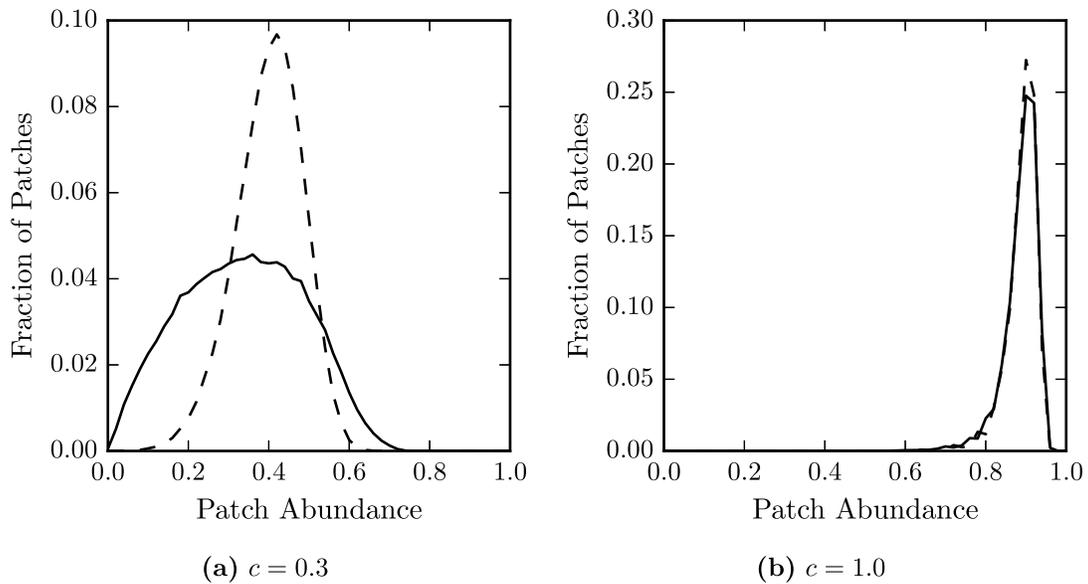}
\caption{ \label{fig3} The distribution of patch abundances for species $A$ at different colonisation rates. The two colonisation rates correspond to different regimes of overall species abundance. At $c=0.3$ overall abundance on \ER networks (dashed lines) is higher than on random geometric networks (solid lines), while at $c=1.0$ overall abundance is similar on both networks.  For \ER networks the shapes of the distributions are similar, but for the random geometric networks the shape of the distribution changes dramatically. a) At $c=1.0$ the distribution on random geometric networks is broad, there are many more patches with lower abundance than in the \ER networks. b) At $c=0.3$  the distribution is similar for both networks. The high overall abundance reduces the impact of spatial separations.}
\label{abundance_distribution}
\end{figure*}

\section{Single Species}

To begin we compare the dispersal of a single species on networks from \ERR{} and \RGG{} ensembles. We find that above the persistence threshold species abundance is higher on \ERR{} networks than \RGG{} networks, see Fig.~\ref{abundance_against_c}. This means that populations of the species exist at more patches in \ERR{} networks than \RGG{} networks. Therefore, the global extinction risk for a single species is comparatively greater on \RGG{} than \ERR{} networks.

The difference in species abundance between the network ensembles is greatest close to the threshold, while for colonisation rates far above the threshold abundance is similar on both ensembles, see Fig.~\ref{abundance_against_c}.  In general the risk of global extinction is higher when abundance is lower. Therefore, the increase in extinction risk to a species on \RGG{}, in comparison to \ERR{} networks, is greatest when the risk of global extinction is greatest.

Let us now try to explain these findings with respect to the spatial structure of \RGG{} networks. Considering parameters near the threshold, we find that the distribution of patch-wise occupations is broader on \RGG{} than \ERR{} networks, see Fig.~\ref{fig3}a. Some patches in \RGG{} networks are occupied for a larger fraction of the time than any patches in \ERR{} networks, while others are occupied for a smaller fraction of the time than any in \ERR{} networks. As extinction is the same at all patches this means that some patches in the \RGG{} networks are, when empty, colonised more quickly by the species than others.

To explain the variation in colonisation between patches we must consider which properties of a patch determine how easy it is to colonise. It is well understood that a patch's degree is correlated with its occupation.\cite{Pastor-Satorras2001,Barter2016} The distribution of patch occupations on \ERR{} networks has a similar shape to the degree distribution, reflecting the influence of degree on patch occupation. The broad distribution of patch occupation in \RGG{} networks reflects the greater prominence non-degree effects in its determination. This indicates that in the large world network the detailed spatial structure, rather than just the overall connectivity, is of greater importance. 

The probability of an unoccupied patch being colonised is affected by the occupation of its neighbours. An unoccupied patch with frequently occupied neighbours will be colonised  quickly, and so be unoccupied for a relatively short period. On the other hand, an unoccupied patch with infrequently occupied neighbours may not be colonised for a longer period. Therefore the properties of the neighbourhood of a patch can determine how often it is occupied.

The observed differences between the distribution of patch occupations in \ERR{} and \RGG{} networks can thus be intuitively explained by the different average distances in these networks. In the \ERR{} networks the short distances mean the neighbourhoods of all patches are similar. In the \RGG{} networks the long distances mean local neighbourhoods can be more varied between nodes.

\begin{figure*}[t]
\centering
\includegraphics{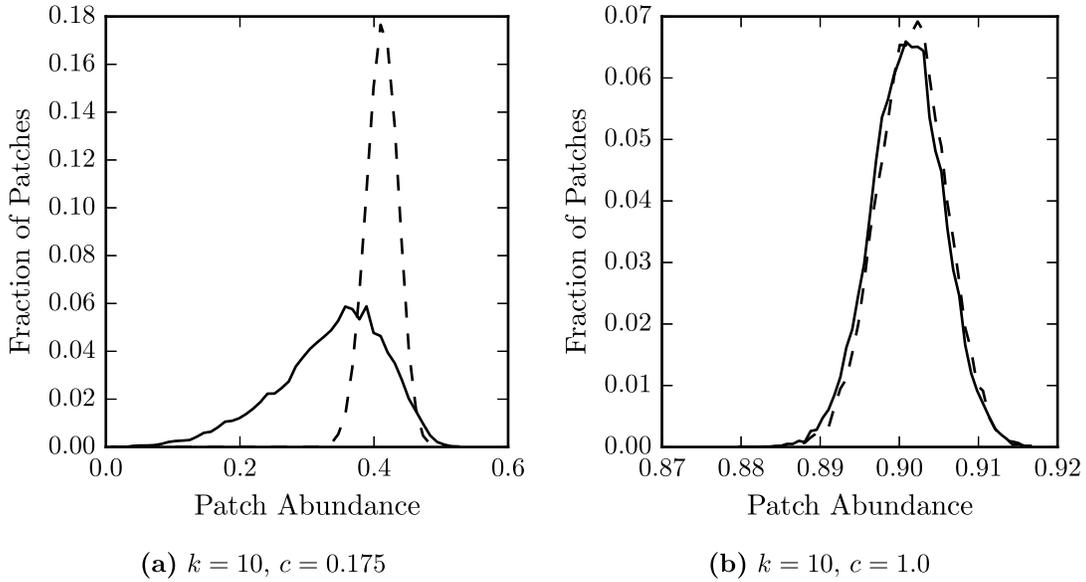}
\caption[Spatial embedding increases variation in occupation of patches with the same degree, while degree correlations increase variation between patches with different degrees]{\label{fig:A_Degs}\label{fig4} 
 Spatial embedding increases variation in occupation of patches with the same degree. The distributions of the fraction of time (patch occupation) that a primary producer occupies patches with  degree $k=10$, at different colonisation rates, $c$, in \ER (dashed line) and random geometric networks (solid line). a) At a relatively low colonisation rate, when total abundance of the species is higher on \ER networks than random geometric networks. b) At a relatively high colonisation rate, when total abundance is similar on both types of network. When overall abundance is low in random geometric networks, patches with the same degree have a broad range of occupations.} 
\end{figure*}

We test the effect of differences between neighbourhoods explicitly by considering the occupation of nodes with only the degree $10$ in Fig.~\ref{fig4}a. We find that the occupation of patches with degree $10$ in \ERR{} networks has a distribution that is narrower than the distribution for all patches. For \RGG{} networks we find that the the occupation of patches with degree $10$ has a relatively broad distribution, only a small amount narrower than the distribution of all patches. Therefore, patches of the same degree have more varied occupations in \RGG{} networks than in \ERR{} networks. This implies that, in spatial networks, the structure of a patch's neighbourhood has a large influence on its occupation by the species. 

\begin{figure*}[t]
\centering 
\includegraphics{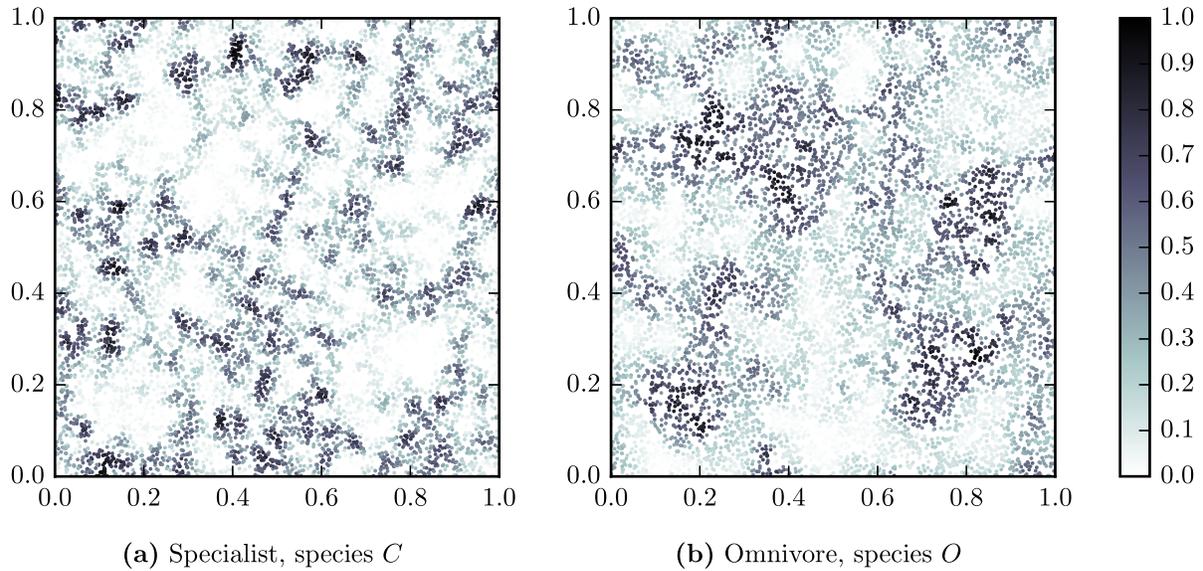}
\caption{\label{fig6}Species abundance on the patches of an example random geometric network with $10^4$ nodes and $\mean{k}=10$ with $c=0.8$. a) Occupation by the specialist is highest in the regions most dense with patches, and the regions close to them. b) Occupation by the omnivore is highest in regions of medium density which are spatially separated from regions with high specialist occupation. The regions with high omnivore occupation complement the regions of high specialist occupation.}
\end{figure*}

By observing the occupation of patches on example networks, such as in Fig.~\ref{fig6}a, we find meso-scale patterns of patch occupation. There are distinct regions of high occupation patches separate from regions of low occupation patches. In the large world network the species is distributed unevenly across the landscape, as regions with different local structures are separated by large spatial distances. In \RGG{} networks there are often more low occupation regions than high occupation regions, and as such the species abundance is lower on these than equivalent \ERR{} networks. Spatial distances reduce the number of patches the species can easily colonise and so abundance is lower in the large world networks.

We now consider the situation at higher values of $c$, for which the species has similar abundance on both types of network. At these parameter values the distribution of patch occupations is similar on both types of network, see Fig.~\ref{fig3}b. Furthermore, the distributions are similar when considering only patches of degree $10$, see Fig.~\ref{fig4}b, suggesting that at these parameter values degree is a good indicator of occupation on \RGG{} networks as well as \ERR{} networks. When overall abundance is high the neighbourhoods of nodes in the large world are more similar and so have a lesser impact on the occupation of individual patches.

Observing example networks shows that spatial variations do exist in networks at high overall abundance, even though degree is a good indicator of patch occupation. This is because the \RGG{} networks have degree correlations. In \RGG{} networks high degree nodes have more high degree neighbours and low degree nodes have more low degree neighbours. Therefore spatial regions with many high degree nodes are more highly occupied that spatial regions with many low degree nodes. In the \RGG{} the degree correlations translate to spatial correlations so the species abundance varies between different regions.

\begin{figure*}[t]
\centering
\includegraphics{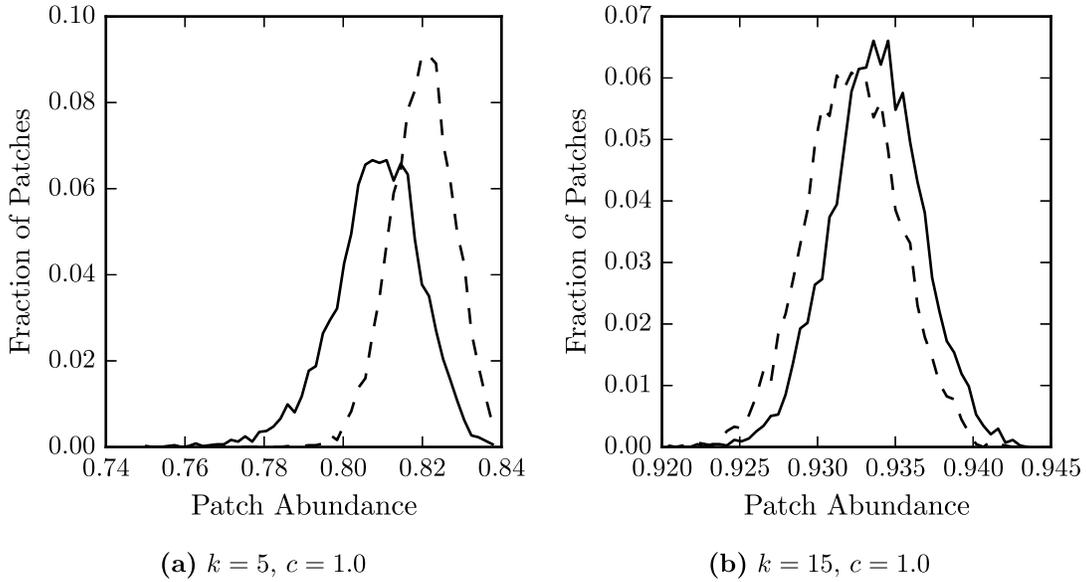}
\caption[Degree correlations increase variation between patches with different degrees]{\label{fig:A_Degs2}\label{fig5} 
Degree correlations increase variation between patches with different degrees. The distributions of the fraction of time (patch occupation) that a primary producer occupies patches with different degrees, $k$, in \ER (dashed line) and random geometric networks (solid line). a) Patches with the relatively low degree, $k=5$ and b) patches with the relatively high degree, $k=15$, both at a colonisation rate, $c=1.0$, when total abundance of the species is similar on both types of network. Degree correlations in random geometric networks lead to lower occupation of patches with low degree and higher occupation of patches with high degree, compared to patches in the \ER networks.} 
\end{figure*}

Furthermore, we find that the mean occupation of patches in \RGG{} networks with relatively low degree is lower than for those in \ERR{} networks,  while the mean occupation of patches in \RGG{} networks with relatively high degrees is higher than for those in \ERR{} networks, see Fig.~\ref{fig:A_Degs2}. The effects of degree on occupation are enhanced on \RGG{} networks due to degree correlations. Typical high degree patches in \RGG{} networks also have high degree neighbours and are colonised more easily than typical high degree patches in \ERR{} networks. Therefore, as well as the regional variations due to large network distances, in spatial networks the species distribution is uneven due to the structural correlations of nearby nodes.

To summarise, both long network distances and degree correlations cause the emergence of meso-scale patterns in the occupation of patches in the random geometric network. At low overall abundance the spatial separations have the dominant effect, and the results is large variations between the occupation of different patches. At high overall abundance  spatial patterns are predominantly due to degree correlations, and these result in smaller spatial variations in patch-wise occupation.

\section{Four species foodweb}

We now extend our study to the meta-foodweb shown in Fig.~\ref{fig1}a, which includes both predator-prey feeding and specialist-omnivore competition interactions.

Species $A$, a primary producer, that can survive on abiotic resources; species $B$, a secondary consumer, that feeds upon the primary producer; species $C$, a top predator, that feeds upon the secondary consumer; and species $O$, an omnivore that preys upon the primary producer and secondary consumer. 

Species $B$, $C$ and $O$ can only occupy patches alongside their prey. In addition, species $O$ is out competed by the specialist predators and therefore cannot survive on patches where species $C$ is established. When both species $B$ and species $O$ occupy a patch, species $B$ out-competes species $O$ for feeding on species $A$, however species $O$ can still survive by feeding on species $B$ instead. 

For each species, $X$, there is a range of values of $c_X/e_X$ for which that species is unable to persist in a network. For specialist species the range is characterized by a critical value of $\mu_X$. When $c_X/e_X<\mu_X$ the species cannot persist in the network and will become globally extinct, but when $c_X/e_X>\mu_X$ the species will persist and be expected to occupy a quasi-steady fraction of the patches. \cite{Bohme2012,Barter2016} For an omnivore species $Y$ which is a competitor to species $X$ there are two critical values. Species $Y$ can only persist in the network when both  $c_Y/e_Y>\mu_Y$, and  $c_X/e_X<\nu_Y$, where the limit $\nu_Y$ is itself dependent on $c_Y/e_Y$.  When $c_X/e_X>\nu_Y$ the omnivore's competitor occupies a large fraction of the patches and the omnivore is unable to persist anywhere in the network, becoming globally extinct. Though we have set $c_X/e_X=c$ for all species, due to indirect extinctions the threshold value, $\mu_X$, is not the same for all species.

Using the foodweb, we can establish how the spatial embedding affects species that interact with feeding or competitive interactions. We investigate: a) how the distribution of a predator is affected by the uneven distribution of its prey, and b) how the distribution of an omnivore is affected by the uneven distribution of its superior competitor.

We start by studying the food chain of species $A$, $B$ and $C$ that contains only predator-prey relationships. We find that all these specialist species have similar behaviour to a single species, and are less abundant in \RGG{} than \ERR{} networks with the same degree distributions under the same dispersal conditions, see Fig.~\ref{abundance_against_c}.

The set of patches occupied by a predator is a subset of the set of patches occupied by its prey. Furthermore the effective extinction rate of the predator is larger than that of its prey. Therefore it is harder for a predator to disperse and the abundance of a predator cannot be greater than the abundance of the prey species it consumes. We find that the abundance of predators is lower on \RGG{} networks even at parameter values where the abundance of their prey is similar in both ensembles.  This suggests that the spatial effects of the underlying network experienced by the single species are experienced by all species in the chain.

\begin{figure*}[t]
\centering
\includegraphics{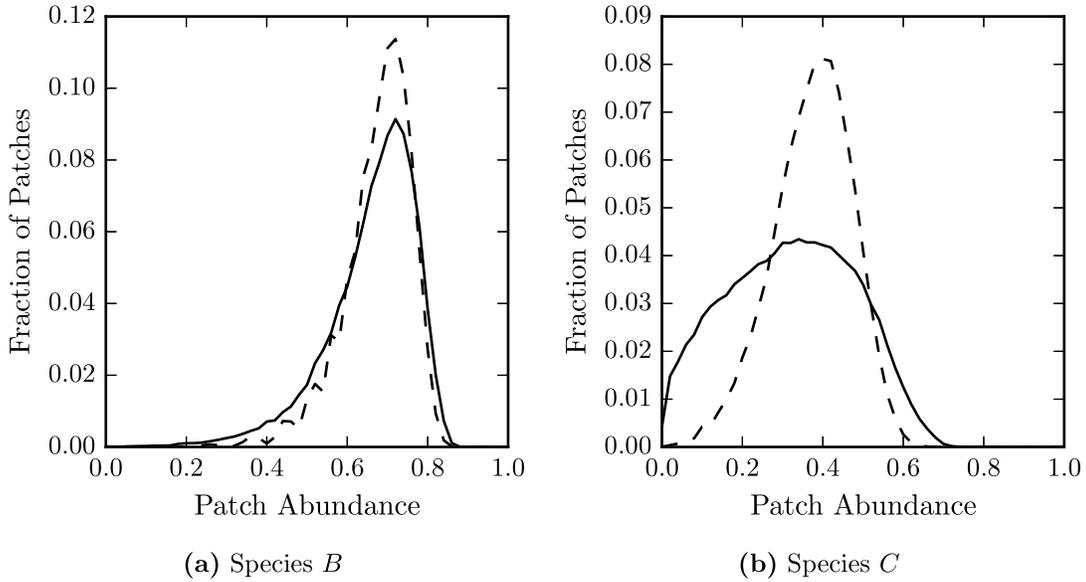}
\caption[Variations in patch occupation are larger for predators than prey]{\label{fig:BC_Patchwise}\label{fig7} Variations in patch occupation are larger for predators than prey. Comparison of the occupation of a) species~$B$ and its predator b) species $C$ on patches from \ER networks (dashed lines) and random geometric networks (solid lines) at $c=1.0$. The variation of patch occupation for the predator, species $C$, on random geometric graphs is larger than for the prey, species $B$, due to the combination of direct spatial effects, and the uneven distribution of the prey.} 
\end{figure*}

For the single species on \RGG{} networks, we found that structural properties lead to spatial variations in patch abundance.  We find that a predator utilising a species as its food source has an even more uneven distribution than its prey, see Fig.~\ref{fig:BC_Patchwise}. The variation in the in habitability of patches in the underlying network is exaggerated by the dispersal of intermediate species in the chain. For species $A$, its prey occupies all patches equally, but structural properties of the network mean it will find some patches are more hospitable than others. For species $B$, preying on species $A$, the structural properties have the same effects, and in addition its prey is unevenly distributed. As species $A$ and $B$ undergo similar dispersal processes the uneven distribution of prey aligns with the structural variations. Therefore, species $B$ experiences greater variation in  habitability, and therefore occupation, across the patches in the network. The local structures in large world networks have a greater impact on the distribution of species higher in the food chain.

\begin{figure*}[t]
\centering
\includegraphics{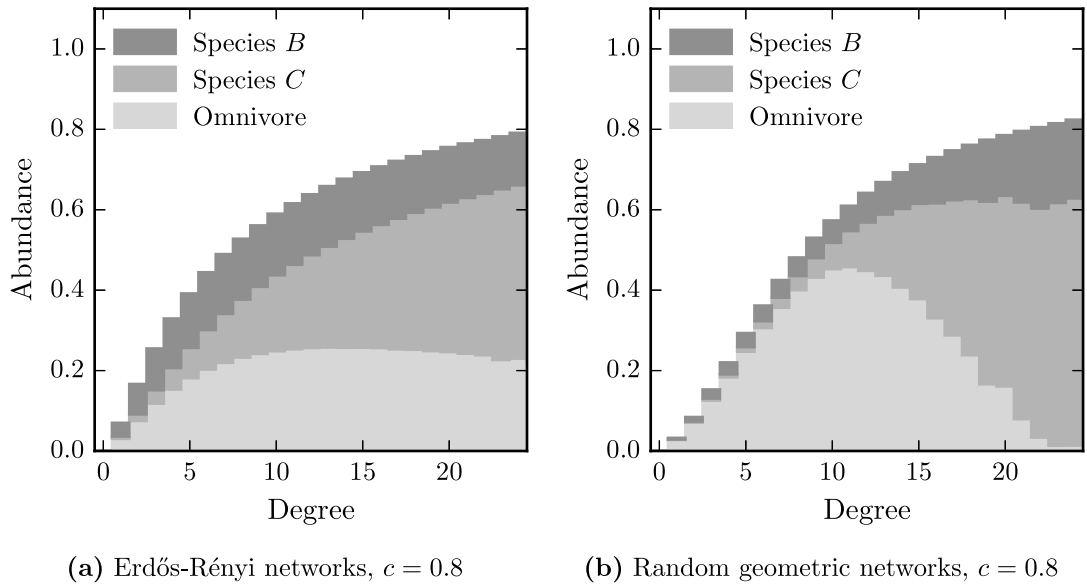}
\caption{\label{fig8}The mean fraction of time patches of different degrees are occupied by each species. The bars are shaded in three sections, these show the fraction of time patches are occupied by: bottom) the omnivore, this can be with or without species $B$; middle) species $C$, which must also be with species $B$; and top) species $B$ but neither species $C$ or the omnivore. In the random geometric networks the intermediate degree patches have low abundance of species $C$, and high abundance of the omnivore.}
\label{degree_fractions}
\end{figure*}

Now let us focus on the effects on the distribution of the omnivore species $O$ due to coexisting with the meta-population of the species $C$. The two species coexist only for colonisation rates in the range $\mu_C<c<\nu_{O}$. When $c<\mu_{C}$, species $C$ cannot persist in the meta community and so the patches available to species $O$ are the same patches available to species $B$, and the omnivore behaves identically to the specialist species $B$. When  $c>\nu_{O}$ the omnivore becomes extinct due to driven extinction from competition with the specialist.

When the specialist and omnivore coexist, the omnivores abundance is greater on \RGG{} networks than on \ERR{} networks, see Fig.~\ref{abundance_against_c}. We previously saw that the large distances in spatial networks hinder the occupation of patches by the specialists, by contrast these distances aid the occupation of patches by the omnivore. 

One reason the omnivore is more abundant on \RGG{} than \ERR{} networks, in the coexistence regime, is the lower overall abundance of species $C$. More of the patches are available to the omnivore for more of the time in the \RGG{} networks. Further, colonisation of a patch that is occupied by species $O$ by species $C$,  causes the local extinction of species $O$. Therefore when a patch which is occupied by species $O$ and species $B$ neighbours a patch occupied by species $C$ the colonisation rate of species $C$ acts to increase the effective extinction rate of species $O$. In spatial networks omnivore populations encounter specialist populations less frequently, and competition has a smaller effect on the occupation of patches by the omnivores.

We have seen that, on both \ERR{} and \RGG{} networks occupation by specialist species increases with node degree, see Fig.~\ref{degree_fractions}. However the variation is  greater in \RGG{} networks, with high degree nodes more likely to be colonised than lower degree nodes. We now investigate whether the greater variation in patch occupation by the specialist on \RGG{} networks, affects the distribution of the omnivore.

On both types of network, at $c=0.8$, occupation by the omnivore is largest on nodes with an intermediate degree. Low degree nodes are hard for all species to occupy, while the large amount of time species $C$ occupies high degree nodes means that they are often unavailable to the omnivore. 

The effect of competitor occupation on omnivore occupation of high degree nodes is  greater on \RGG{} than on \ERR{} networks. Figure~\ref{fig8}b for \RGG{} networks, has a pronounced peak in omnivore occupation at intermediate degrees. The high occupation by the specialist of high degree patches makes them almost always unavailable to the omnivore and the omnivore rarely occupies them. In spatial networks some patches are occupied by the specialist so often that a population of omnivores is almost never established on them.

Patches with intermediate degrees are relatively unlikely to be occupied by the specialist on \RGG{} networks compared to \ERR{} networks. Subsequently, the occupation of these patches by the omnivore is high, almost as high as occupation by the omnivore in the absence of competition, which is equivalent to occupation by species $B$. The large network distances mean that the omnivore is able to colonise these nodes and establish populations that rarely encounter competition from specialist populations.

Both \RGG{} and \ERR{} networks have poisson degree distributions, and many more patches with intermediate degree than with the highest degrees. Subsequently, the overall abundance of the omnivore is greater on the \RGG{} than the \ERR{} networks. The spatial distances in large worlds make more patches more hospitable to the omnivore than they make less hospitable, and therefore decrease the chance of global omnivore extinction.

To summarise, the large path lengths in \RGG{} networks mean that some regions of the graph are hospitable to the omnivore but inhospitable to the specialist. Therefore, the omnivore can colonise these regions without suffering regular competitive extinctions, and consequentially, occupy them a large fraction of the time. In other words, the omnivore can fill in gaps in the specialists dispersal, see Fig.~\ref{fig6}b.

\section{Conclusions and discussion}

We used agent-based simulations to investigate the dispersal of a foodweb, comprising predator-prey and competitive interactions, on spatially embedded patch networks.

The abundance of primary producers and specialist consumers, which are only affected by predator-prey interactions, is lower on spatially embedded random geometric patch networks than non-spatial \ER patch networks. Furthermore, in spatially embedded networks there is a greater variation between the fraction of time different patches are occupied by any species. By visualising patch abundance of specialists, we found that these variations correspond meso-scale patters of patch occupation. Regions dense with patches of high occupation are spatially separated from regions of patches with low occupation.  

A specialist consumer only occupies patches that its prey also occupies. In general its prey is also a specialist consumer. The local factors, such as high degree, that make a patch easy for the prey species to occupy it also make it easy for its predator to occupy. Further, spatial variations in prey occupations reduce the availability of patches for its predator in some regions of spatial networks. Consequently, the difference in the abundance of a predator between \ER and random geometric networks is larger than the difference in the abundance of its prey.

We assume that generalist species, such as omnivores, are weaker competitors than specialist species with the same prey¸ such that the generalist is driven to local extinction in any interaction. Subsequently, reduced overall specialist abundance in random geometric networks aids generalist persistence. Further, we identify that the meso-scale structure in spatially embedded networks aids the persistence of generalist species in three ways. Firstly, patches can have relatively high degrees but low specialist abundance due to network separation from the most occupied regions of the network and patches with higher degree are easier for the generalist to colonise. Secondly, patches with low specialist occupation are often grouped together and the generalist can persist in such a region without stochastic extinction. Thirdly, regions with low specialist occupation are separated by long distances from the regions of high specialist occupation and so the generalist populations there are rarely threatened by driven extinction. The combination of these factors leads to greater abundance of generalist species on random geometric networks, compared to \ER networks.

Two properties of the spatially embedded networks cause the significant regional variations in patch occupation: long distances between nodes and degree correlations. Previous work has indicated that long distances have a greater effect than degree correlations on dispersal processes,\cite{Isham2011} and our results also support this conclusion. When regional variations are large, degree is a poor indicator of patch occupation, but when regional variations are small it is a good indicator. This suggests that when regional variations are larger, they are being primarily caused by long network distances and when regional variations are smaller we are seeing the effect only of degree correlations. In other words, long distances have a greater effect than degree correlations on species dispersal over spatial networks.

Real patch landscapes are usually inherently spatial and these results suggest that structures resulting from this have many implications for environmental management. For example, our results demonstrate the contrasting results possible from measuring diversity at different spatial scales.\cite{Whittaker1972}  Further, the ability of a species to inhabit a patch is dependent on the structure of surrounding patches and long range effects. In situations, such as deforestation, where a managed removal of patches attempts to preserve habitat by preserving a single, well populated patch, these attempts may fail if the neighbourhood of that patch is destroyed such that recolonisation becomes more difficult. Alternatively, well intentioned environmental management to maintain a specialist species can destroy the spatial separation that allows omnivores to persist, sheltered from specialist populations. 

The spatial embedding of agents is a property of many networks found in ecology \cite{pilosof2015ecological} and elsewhere. \cite{Riley2015} However, dynamic process on networks incorporating spatial structure are relatively poorly understood. Most existing analytical approaches for dispersal processes on networks focus on local structure,\cite{Pastor-Satorras2015}  such as degree correlations. Our results show that local properties in spatial networks often have a smaller impact than long network distances. Though there has been some recent development of methods for considering spatial separation of nodes in networks,\cite{Estrada2016} further advances are required to characterise space in network models, and understand its effects on dynamic processes.

One particular challenge for analytical approaches to spatial networks is how to characterise their spatial structure. The most popular model for generating spatial networks, using random geometric graphs, results in networks with several notable properties that make them different from random networks. For instance higher clustering coefficients, larger diameters, larger average path length and positive degree correlations.\cite{Penrose2003} These properties are not independent but are jointly caused by the spatial embedding. For example, nodes near each other are neighbours and the radius containing their neighbourhoods largely intersect, resulting in clustering and also degree correlations. Our results show that both long separations and local properties are simultaneously important in explaining the dynamical behaviour of systems on spatial networks. Hence,  methods to better understand these systems should be capable of accounting for the combination of these properties.

We finish by suggesting how this may be accomplished. It may be possible to approximate the regions of a spatial graph of homogeneous patches by a graph of heterogeneous meta-patches. The properties of each of these meta-patches would be constructed to reflect the region of the underlying spatial graph that the particular patch represents. A random geometric graph in this model would have meta-patches with properties drawn from a distribution determined by the formation processes of a random geometric graph. For example, each meta-patch could have a local extinction probability, based on the number of patches included and the number of links between them. The graph of meta-patches contains many fewer patches, and could be potentially be analysed using available numerical methods. The sensitivity of the model to meta-patch properties may be analysed to determine the systems sensitivity to the underlying properties of spatial patch networks that inform them. 

\newpage

\section*{Data Availability}
This study did not involve any underlying data.

\bibliography{RGG1.bib}

\begin{thebibliography}{10}
\expandafter\ifx\csname url\endcsname\relax
  \def\url#1{\texttt{#1}}\fi
\expandafter\ifx\csname urlprefix\endcsname\relax\def\urlprefix{URL }\fi
\providecommand{\bibinfo}[2]{#2}
\providecommand{\eprint}[2][]{\url{#2}}

\bibitem{May1972}
\bibinfo{author}{May, R.~M.}
\newblock \bibinfo{title}{{Will a Large Complex System be Stable?}}
\newblock \emph{\bibinfo{journal}{Nature}} \textbf{\bibinfo{volume}{238}},
  \bibinfo{pages}{413--414} (\bibinfo{year}{1972}).

\bibitem{McCann1998}
\bibinfo{author}{McCann, K.}, \bibinfo{author}{Hastings, A.} \&
  \bibinfo{author}{Huxel, G.~R.}
\newblock \bibinfo{title}{{Weak trophic interactions and the balance of
  nature}}.
\newblock \emph{\bibinfo{journal}{Nature}} \textbf{\bibinfo{volume}{395}},
  \bibinfo{pages}{794--798} (\bibinfo{year}{1998}).

\bibitem{McCann2000}
\bibinfo{author}{McCann, K.~S.}
\newblock \bibinfo{title}{{The diversity-stability debate}}.
\newblock \emph{\bibinfo{journal}{Nature}} \textbf{\bibinfo{volume}{405}},
  \bibinfo{pages}{228--233} (\bibinfo{year}{2000}).

\bibitem{Neutel2002}
\bibinfo{author}{Neutel, A.-M.}, \bibinfo{author}{Heesterbeek, J. A.~P.} \&
  \bibinfo{author}{de~Ruiter, P.~C.}
\newblock \bibinfo{title}{{Stability in Real Food Webs: Weak Links in Long
  Loops}}.
\newblock \emph{\bibinfo{journal}{Science}} \textbf{\bibinfo{volume}{296}},
  \bibinfo{pages}{1120--1123} (\bibinfo{year}{2002}).

\bibitem{Montoya2006}
\bibinfo{author}{Montoya, J.~M.}, \bibinfo{author}{Pimm, S.~L.} \&
  \bibinfo{author}{Sol{\'{e}}, R.~V.}
\newblock \bibinfo{title}{{Ecological networks and their fragility}}.
\newblock \emph{\bibinfo{journal}{Nature}} \textbf{\bibinfo{volume}{442}},
  \bibinfo{pages}{259--264} (\bibinfo{year}{2006}).

\bibitem{Gross2009}
\bibinfo{author}{Gross, T.}, \bibinfo{author}{Rudolf, L.},
  \bibinfo{author}{Levin, S.~A.} \& \bibinfo{author}{Dieckmann, U.}
\newblock \bibinfo{title}{{Generalized models reveal stabilizing factors in
  food webs.}}
\newblock \emph{\bibinfo{journal}{Science (New York, N.Y.)}}
  \textbf{\bibinfo{volume}{325}}, \bibinfo{pages}{747--50}
  (\bibinfo{year}{2009}).

\bibitem{Lafferty2015}
\bibinfo{author}{Lafferty, K.~D.} \emph{et~al.}
\newblock \bibinfo{title}{{A general consumer-resource population model}}.
\newblock \emph{\bibinfo{journal}{Science}} \textbf{\bibinfo{volume}{349}},
  \bibinfo{pages}{854--857} (\bibinfo{year}{2015}).

\bibitem{Levins1969}
\bibinfo{author}{Levins, R.}
\newblock \bibinfo{title}{{Some Demographic and Genetic Consequences of
  Environmental Heterogeneity for Biological Control}}.
\newblock \emph{\bibinfo{journal}{Bulletin of the Entomological Society of
  America}} \textbf{\bibinfo{volume}{15}}, \bibinfo{pages}{237--240}
  (\bibinfo{year}{1969}).

\bibitem{Hanski1999}
\bibinfo{author}{Hanski, I.}
\newblock \emph{\bibinfo{title}{{Metapopulation Ecology}}}
  (\bibinfo{publisher}{Oxford University Press}, \bibinfo{year}{1999}).

\bibitem{Warfe2013}
\bibinfo{author}{Warfe, D.~M.} \emph{et~al.}
\newblock \bibinfo{title}{Productivity, disturbance and ecosystem size have no
  influence on food chain length in seasonally connected rivers}.
\newblock \emph{\bibinfo{journal}{PLOS ONE}} \textbf{\bibinfo{volume}{8}},
  \bibinfo{pages}{1--11} (\bibinfo{year}{2013}).

\bibitem{Gramlich2016}
\bibinfo{author}{Gramlich, P.}, \bibinfo{author}{Plitzko, S.},
  \bibinfo{author}{Rudolf, L.}, \bibinfo{author}{Drossel, B.} \&
  \bibinfo{author}{Gross, T.}
\newblock \bibinfo{title}{The influence of dispersal on a predator–prey
  system with two habitats}.
\newblock \emph{\bibinfo{journal}{Journal of Theoretical Biology}}
  \textbf{\bibinfo{volume}{398}}, \bibinfo{pages}{150 -- 161}
  (\bibinfo{year}{2016}).

\bibitem{Kivela2014}
\bibinfo{author}{Kivela, M.} \emph{et~al.}
\newblock \bibinfo{title}{{Multilayer networks}}.
\newblock \emph{\bibinfo{journal}{Journal of Complex Networks}}
  \textbf{\bibinfo{volume}{2}}, \bibinfo{pages}{203--271}
  (\bibinfo{year}{2014}).

\bibitem{MacArthur1967}
\bibinfo{author}{MacArthur, R.~H.} \& \bibinfo{author}{Wilson, E.~O.}
\newblock \emph{\bibinfo{title}{{Theory of Island Biogeography}}}
  (\bibinfo{publisher}{Princeton University Press}, \bibinfo{year}{1967}).

\bibitem{Harrison1991}
\bibinfo{author}{Harrison, S.}
\newblock \bibinfo{title}{{Local extinction in a metapopulation context: an
  empirical evaluation}}.
\newblock \emph{\bibinfo{journal}{Biological journal of the Linnean Society}}
  \textbf{\bibinfo{volume}{42}}, \bibinfo{pages}{73--88}
  (\bibinfo{year}{1991}).

\bibitem{Levins1971}
\bibinfo{author}{Levins, R.} \& \bibinfo{author}{Culver, D.}
\newblock \bibinfo{title}{{Regional Coexistence of Species and Competition
  between Rare Species}} \textbf{\bibinfo{volume}{68}},
  \bibinfo{pages}{1246--1248} (\bibinfo{year}{1971}).

\bibitem{Levin1974}
\bibinfo{author}{Levin, S.~A.}
\newblock \bibinfo{title}{{Dispersion and Population Interactions}}.
\newblock \emph{\bibinfo{journal}{The American Naturalist}}
  \textbf{\bibinfo{volume}{108}}, \bibinfo{pages}{207--228}
  (\bibinfo{year}{1974}).

\bibitem{Nee1992}
\bibinfo{author}{Nee, S.} \& \bibinfo{author}{May, R.~M.}
\newblock \bibinfo{title}{{Dynamics of Metapopulations: Habitat Destruction and
  Competitive Coexistence}}.
\newblock \emph{\bibinfo{journal}{Journal of Animal Ecology}}
  \textbf{\bibinfo{volume}{61}}, \bibinfo{pages}{37--40}
  (\bibinfo{year}{1992}).

\bibitem{Hanski1991}
\bibinfo{author}{Hanski, I.} \& \bibinfo{author}{Gilpin, M.}
\newblock \bibinfo{title}{{Metapopulation dynamics: brief history and
  conceptual domain}}.
\newblock \emph{\bibinfo{journal}{Biological Journal of the Linnean Society}}
  \textbf{\bibinfo{volume}{42}}, \bibinfo{pages}{3--16} (\bibinfo{year}{1991}).

\bibitem{Hanski1998}
\bibinfo{author}{Hanski, I.}
\newblock \bibinfo{title}{{Metapopulation dynamics}}.
\newblock \emph{\bibinfo{journal}{Nature}} \textbf{\bibinfo{volume}{396}},
  \bibinfo{pages}{41--49} (\bibinfo{year}{1998}).

\bibitem{Hanski1999a}
\bibinfo{author}{Hanski, I.}
\newblock \bibinfo{title}{{The Levins model and its variants}}.
\newblock In \emph{\bibinfo{booktitle}{Metapopulation Ecology}},
  chap.~\bibinfo{chapter}{4}, \bibinfo{pages}{55--75}
  (\bibinfo{publisher}{Oxford University Press}, \bibinfo{year}{1999}).

\bibitem{Pillai2009}
\bibinfo{author}{Pillai, P.}, \bibinfo{author}{Loreau, M.} \&
  \bibinfo{author}{Gonzalez, A.}
\newblock \bibinfo{title}{{A patch-dynamic framework for food web
  metacommunities}}.
\newblock \emph{\bibinfo{journal}{Theoretical Ecology}}
  \textbf{\bibinfo{volume}{3}}, \bibinfo{pages}{223--237}
  (\bibinfo{year}{2009}).

\bibitem{Pillai2011}
\bibinfo{author}{Pillai, P.}, \bibinfo{author}{Gonzalez, A.} \&
  \bibinfo{author}{Loreau, M.}
\newblock \bibinfo{title}{{Metacommunity theory explains the emergence of food
  web complexity.}}
\newblock \emph{\bibinfo{journal}{Proceedings of the National Academy of
  Sciences of the United States of America}} \textbf{\bibinfo{volume}{108}},
  \bibinfo{pages}{19293--8} (\bibinfo{year}{2011}).

\bibitem{Gravel2011}
\bibinfo{author}{Gravel, D.}, \bibinfo{author}{Canard, E.},
  \bibinfo{author}{Guichard, F.} \& \bibinfo{author}{Mouquet, N.}
\newblock \bibinfo{title}{{Persistence increases with diversity and connectance
  in trophic metacommunities.}}
\newblock \emph{\bibinfo{journal}{PloS one}} \textbf{\bibinfo{volume}{6}},
  \bibinfo{pages}{e19374} (\bibinfo{year}{2011}).

\bibitem{Bohme2012}
\bibinfo{author}{B{\"{o}}hme, G.~A.} \& \bibinfo{author}{Gross, T.}
\newblock \bibinfo{title}{{Persistence of complex food webs in
  metacommunities}}.
\newblock \emph{\bibinfo{journal}{arXiv:1212.5025}}  (\bibinfo{year}{2012}).

\bibitem{Barter2016}
\bibinfo{author}{Barter, E.} \& \bibinfo{author}{Gross, T.}
\newblock \bibinfo{title}{{Meta-food-chains as a many-layer epidemic process on
  networks}}.
\newblock \emph{\bibinfo{journal}{Phys. Rev. E}} \textbf{\bibinfo{volume}{93}},
  \bibinfo{pages}{22303} (\bibinfo{year}{2016}).

\bibitem{Newman2010}
\bibinfo{author}{Newman, M.}
\newblock \emph{\bibinfo{title}{{Networks: An Introduction}}}
  (\bibinfo{publisher}{Oxford University Press}, \bibinfo{year}{2010}).

\bibitem{Klee1981}
\bibinfo{author}{Klee, V.} \& \bibinfo{author}{Larman, D.}
\newblock \bibinfo{title}{{Diameters of random graphs}}.
\newblock \emph{\bibinfo{journal}{Canad. J. Math}}
  \textbf{\bibinfo{volume}{33}}, \bibinfo{pages}{618--640}
  (\bibinfo{year}{1981}).

\bibitem{Bollobas1981}
\bibinfo{author}{Bollob{\'{a}}s, B.}
\newblock \bibinfo{title}{{The Diameter of Random Graphs}}.
\newblock \emph{\bibinfo{journal}{Transactions of the American Mathematical
  Society}} \textbf{\bibinfo{volume}{267}}, \bibinfo{pages}{41--52}
  (\bibinfo{year}{1981}).

\bibitem{Watts1998}
\bibinfo{author}{Watts, D.~J.} \& \bibinfo{author}{Strogatz, S.~H.}
\newblock \bibinfo{title}{{Collective dynamics of 'small-world' networks.}}
\newblock \emph{\bibinfo{journal}{Nature}} \textbf{\bibinfo{volume}{393}},
  \bibinfo{pages}{440--442} (\bibinfo{year}{1998}).

\bibitem{Banavar1999}
\bibinfo{author}{Banavar, J.~R.}, \bibinfo{author}{Maritan, A.} \&
  \bibinfo{author}{Rinaldo, A.}
\newblock \bibinfo{title}{{Size and form in efficient transportation
  networks}}.
\newblock \emph{\bibinfo{journal}{Nature}} \textbf{\bibinfo{volume}{399}},
  \bibinfo{pages}{130--132} (\bibinfo{year}{1999}).

\bibitem{Dall2002}
\bibinfo{author}{Dall, J.} \& \bibinfo{author}{Christensen, M.}
\newblock \bibinfo{title}{{Random geometric graphs.}}
\newblock \emph{\bibinfo{journal}{Physical review. E, Statistical, nonlinear,
  and soft matter physics}} \textbf{\bibinfo{volume}{66}},
  \bibinfo{pages}{016121} (\bibinfo{year}{2002}).

\bibitem{erdds1959random}
\bibinfo{author}{Erdős, P.} \& \bibinfo{author}{R{\'{e}}nyi, A.}
\newblock \bibinfo{title}{{On random graphs I.}}
\newblock \emph{\bibinfo{journal}{Publ. Math. Debrecen}}
  \textbf{\bibinfo{volume}{6}}, \bibinfo{pages}{290--297}
  (\bibinfo{year}{1959}).

\bibitem{Gilbert1959}
\bibinfo{author}{Gilbert, A.~N.}
\newblock \bibinfo{title}{{Random Graphs}}.
\newblock \emph{\bibinfo{journal}{Annals of Mathematical statistics}}
  \textbf{\bibinfo{volume}{30}}, \bibinfo{pages}{1141--1144}
  (\bibinfo{year}{1959}).

\bibitem{erdos1961evolution}
\bibinfo{author}{Erdős, P.} \& \bibinfo{author}{R{\'{e}}nyi, A.}
\newblock \bibinfo{title}{{On the evolution of random graphs}}.
\newblock \emph{\bibinfo{journal}{Bull. Inst. Internat. Statist}}
  \textbf{\bibinfo{volume}{38}}, \bibinfo{pages}{343--347}
  (\bibinfo{year}{1961}).

\bibitem{Pastor-Satorras2001}
\bibinfo{author}{Pastor-Satorras, R.} \& \bibinfo{author}{Vespignani, A.}
\newblock \bibinfo{title}{{Epidemic spreading in scale-free networks}}.
\newblock \emph{\bibinfo{journal}{Physical Review Letters}}
  \textbf{\bibinfo{volume}{86}}, \bibinfo{pages}{3200--3203}
  (\bibinfo{year}{2001}).

\bibitem{Penrose2003}
\bibinfo{author}{Penrose, M.}
\newblock \emph{\bibinfo{title}{{Random Geometric Graphs}}}
  (\bibinfo{publisher}{Oxford University Press}, \bibinfo{year}{2003}),
  \bibinfo{edition}{1} edn.

\bibitem{Friedrich2013}
\bibinfo{author}{Friedrich, T.}, \bibinfo{author}{Sauerwald, T.} \&
  \bibinfo{author}{Stauffer, A.}
\newblock \bibinfo{title}{{Diameter and Broadcast Time of Random Geometric
  Graphs in Arbitrary Dimensions}}.
\newblock \emph{\bibinfo{journal}{Algorithmica}} \textbf{\bibinfo{volume}{67}},
  \bibinfo{pages}{65--88} (\bibinfo{year}{2013}).

\bibitem{Gillespie1976}
\bibinfo{author}{Gillespie, D.~T.}
\newblock \bibinfo{title}{{A general method for numerically simulating the
  stochastic time evolution of coupled chemical reactions}}.
\newblock \emph{\bibinfo{journal}{Journal of Computational Physics}}
  \textbf{\bibinfo{volume}{22}}, \bibinfo{pages}{403--434}
  (\bibinfo{year}{1976}).

\bibitem{Isham2011}
\bibinfo{author}{Isham, V.}, \bibinfo{author}{Kaczmarska, J.} \&
  \bibinfo{author}{Nekovee, M.}
\newblock \bibinfo{title}{Spread of information and infection on finite random
  networks}.
\newblock \emph{\bibinfo{journal}{Phys. Rev. E}} \textbf{\bibinfo{volume}{83}},
  \bibinfo{pages}{046128} (\bibinfo{year}{2011}).

\bibitem{Whittaker1972}
\bibinfo{author}{Whittaker, R.~H.}
\newblock \bibinfo{title}{Evolution and measurement of species diversity}.
\newblock \emph{\bibinfo{journal}{Taxon}} \textbf{\bibinfo{volume}{21}},
  \bibinfo{pages}{213--251} (\bibinfo{year}{1972}).

\bibitem{pilosof2015ecological}
\bibinfo{author}{Pilosof, S.}, \bibinfo{author}{Porter, M.~A.},
  \bibinfo{author}{Pascual, M.} \& \bibinfo{author}{K{\'e}fi, S.}
\newblock \bibinfo{title}{The multilayer nature of ecological networks}.
\newblock \emph{\bibinfo{journal}{Nature Ecology \& Evolution}}
  \textbf{\bibinfo{volume}{1}}, \bibinfo{pages}{0101} (\bibinfo{year}{2017}).

\bibitem{Riley2015}
\bibinfo{author}{Riley, S.}, \bibinfo{author}{Eames, K.},
  \bibinfo{author}{Isham, V.}, \bibinfo{author}{Mollison, D.} \&
  \bibinfo{author}{Trapman, P.}
\newblock \bibinfo{title}{Five challenges for spatial epidemic models}.
\newblock \emph{\bibinfo{journal}{Epidemics}} \textbf{\bibinfo{volume}{10}},
  \bibinfo{pages}{68 -- 71} (\bibinfo{year}{2015}).
\newblock \bibinfo{note}{Challenges in Modelling Infectious \{DIsease\}
  Dynamics}.

\bibitem{Pastor-Satorras2015}
\bibinfo{author}{Pastor-satorras, R.} \emph{et~al.}
\newblock \bibinfo{title}{{Epidemic processes in complex networks}}.
\newblock \emph{\bibinfo{journal}{Reviews of Modern Physics}}
  \textbf{\bibinfo{volume}{87}}, \bibinfo{pages}{1--62} (\bibinfo{year}{2015}).

\bibitem{Estrada2016}
\bibinfo{author}{Estrada, E.}, \bibinfo{author}{Meloni, S.},
  \bibinfo{author}{Sheerin, M.} \& \bibinfo{author}{Moreno, Y.}
\newblock \bibinfo{title}{Epidemic spreading in random rectangular networks}.
\newblock \emph{\bibinfo{journal}{Phys. Rev. E}} \textbf{\bibinfo{volume}{94}},
  \bibinfo{pages}{052316} (\bibinfo{year}{2016}).

\end{thebibliography}
\section*{Acknowledgements}
This work was supported by the EPSRC under grant codes EP/N034384/1 and EP/I013717/1.

\section*{Additional information }
There is no competing interest for any of the authors.

\section*{Author contribution}
All authors contributed to and reviewed the manuscript. E.B. prepared the figures.

\end{document}